\documentclass[conference]{IEEEtran}

\newif\ifshowcomments
\newif\ifshowauthors

\showcommentsfalse

 \showauthorstrue

\usepackage[T1]{fontenc} %
\usepackage[utf8]{inputenc} %
\usepackage[english]{babel} %
\usepackage{amsmath}
\usepackage{amsthm}
\usepackage{amsfonts}
\usepackage{colonequals} %
\usepackage[cmintegrals]{newtxmath} %
\usepackage{mathtools} %
\usepackage{thmtools} %
\usepackage{textcomp}
\usepackage{stmaryrd} %
\usepackage{bm} %
\usepackage{siunitx} %
\usepackage{comment} %
\usepackage{microtype} %
\usepackage{fnpct} %
\usepackage{fontawesome5} %
\usepackage[autostyle=true]{csquotes} %
\usepackage[all]{nowidow} %
\usepackage{balance}
\usepackage[inline]{enumitem} %
\usepackage{tabularx} %
\usepackage[table]{xcolor} %
\usepackage{array}
\usepackage{booktabs} %
\usepackage[hidelinks]{hyperref} %
\usepackage[capitalize,nameinlink]{cleveref} %
\usepackage{graphicx} %
\usepackage[inkscapelatex=false]{svg} %
\usepackage{wrapfig} %
\usepackage{float}  %
\usepackage{placeins} %
\usepackage{xspace} %
\usepackage[noEnd]{algpseudocodex} %
\usepackage[
    backend=biber,
    style=ieee,
    sortcites=true,
    url=true,
    eprint=true,
    giveninits=true,
    minnames=1,
    maxnames=5
]{biblatex}
\AtEveryBibitem{%
    \clearname{editor}%
    \clearfield{series}%
    \clearfield{isbn}%
    \clearfield{issn}%
    \clearfield{day}%
    \clearfield{month}%
    \clearfield{place}%
    \clearlist{location}%
    \clearfield{pages}%
    \clearfield{volume}%
    \clearfield{number}%
    \ifentrytype{software}{%
    }{%
        \clearfield{url}%
        \clearfield{urlyear}%
        \clearfield{urldate}%
    }%
}
\DeclareSourcemap{
    \maps[datatype=bibtex]{
        \map{
            \step[fieldsource=doi, match=\regexp{.+/arXiv\..+}, final]
            \step[fieldsource=eprint, match=\regexp{.+}, final]
            \step[fieldset=doi, null]
        }
        \map{
            \step[fieldsource=doi, match=\regexp{.+}, final]
            \step[fieldset=eprint, null]
            \step[fieldset=archiveprefix, null]
            \step[fieldset=eprinttype, null]
        }
    }
}
\setlist[enumerate]{label=(\arabic*)} %
\makeatletter
\long\def\@makecaption#1#2{%
    \ifx\@captype\@IEEEtablestring%
    \footnotesize\bgroup\par\centering\@IEEEtabletopskipstrut{\normalfont\footnotesize {#1.}\nobreakspace\scshape #2}\par\addvspace{0.5\baselineskip}\egroup%
    \@IEEEtablecaptionsepspace
    \else
        \@IEEEfigurecaptionsepspace
        \setbox\@tempboxa\hbox{\normalfont\footnotesize {#1.}\nobreakspace #2}%
        \ifdim \wd\@tempboxa >\hsize%
        \setbox\@tempboxa\hbox{\normalfont\footnotesize {#1.}\nobreakspace}%
        \parbox[t]{\hsize}{\normalfont\footnotesize \noindent\unhbox\@tempboxa#2}%
        \else%
            \hbox to\hsize{\normalfont\footnotesize\hfil\box\@tempboxa\hfil}%
        \fi\fi}
\makeatother
\makeatletter
\let\MYcaption\@makecaption
\makeatother
\usepackage{subcaption} %
\makeatletter
\let\@makecaption\MYcaption
\makeatother
\graphicspath{ {figs/} }
\svgpath{ {figs/} }
\faStyle{regular}
\declaretheoremstyle[
    bodyfont=\itshape,%
]{example-style}
\declaretheorem[
    name=Example,%
    style=example-style,%
    numbered=unless unique,%
]{example}

\crefname{section}{Sec.}{Sec.}
\Crefname{section}{Sec.}{Sec.}
\crefname{example}{Ex.}{Ex.}
\Crefname{example}{Ex.}{Ex.}
\newcolumntype{R}{>{\raggedleft\arraybackslash}X}
\newcolumntype{C}{>{\centering\arraybackslash}X}
\newcommand{\emptylines}[1]{%
    \newcount\foo%
    \foo=0%
    \loop%
    \phantom{}\\%
    \advance\foo 1%
    \ifnum \foo<#1%
    \repeat%
}

\definecolor{TUM_blue}{RGB}{0,101,189}
\colorlet{TUM_black}{black}
\colorlet{TUM_white}{white}
\definecolor{TUM_darkblue}{RGB}{0,82,147}
\colorlet{TUM_darkblue100}{TUM_darkblue}
\colorlet{TUM_darkblue80}{TUM_darkblue100!80}
\colorlet{TUM_darkblue50}{TUM_darkblue100!50}
\colorlet{TUM_darkblue20}{TUM_darkblue100!20}
\definecolor{TUM_verydarkblue}{RGB}{0,51,89}
\colorlet{TUM_verydarkblue100}{TUM_verydarkblue}
\colorlet{TUM_verydarkblue80}{TUM_verydarkblue100!80}
\colorlet{TUM_verydarkblue50}{TUM_verydarkblue100!50}
\colorlet{TUM_verydarkblue20}{TUM_verydarkblue100!20}
\colorlet{TUM_darkgrey}{TUM_black!80}
\colorlet{TUM_grey}{TUM_black!50}
\colorlet{TUM_lightgrey}{TUM_black!20}
\definecolor{TUM_beige}{RGB}{218,215,203}
\definecolor{TUM_orange}{RGB}{227,114,34}
\definecolor{TUM_green}{RGB}{162,173,0}
\definecolor{TUM_verylightblue}{RGB}{152,198,234}
\definecolor{TUM_lightblue}{RGB}{100,160,200}
\definecolor{IQM_lightgreen}{RGB}{228,245,238}
\definecolor{IQM_darkgreen}{RGB}{76,155,119}
\definecolor{IQM_verydarkgreen}{RGB}{48,88,76}
\definecolor{IQM_lightbrown}{RGB}{241,238,232}
\definecolor{IQM_brown}{RGB}{163,168,162}
\definecolor{IQM_verydarkbrown}{RGB}{70,70,67}

\newcommand{\eg}{e.\,g.\nolinebreak\@\xspace}

\renewcommand{\phi}{\varphi}

\ifshowcomments
\usepackage[textwidth=5cm]{todonotes} %
\usepackage[switch]{lineno} %
\linenumbers
\newlength{\widen}
\advance\paperwidth by \dimexpr\widen+\widen\relax
\advance\evensidemargin by \widen
\advance\oddsidemargin by \widen
\marginparwidth=\dimexpr\widen+0mm\relax %
\newcommand{\mytodo}[4]{\bgroup\color{#2!65!black}#3\egroup\todo[size=\scriptsize,color=#2]{#1: #4}}
\else
\newcommand{\mytodo}[4]{#3}
\usepackage{flushend} %
\fi
\definecolor{marcel}{HTML}{e89c1c}
\definecolor{lukas}{HTML}{e5dd0c}
\definecolor{patrick}{HTML}{a5de0b}
\definecolor{iqm}{HTML}{69DED6}
\definecolor{eric}{HTML}{3fb86e}

\usepackage{pgfplots}
\pgfplotsset{compat=1.18}
\usetikzlibrary{positioning, fit, arrows.meta, calc, backgrounds}

\usepackage{listings}

\lstdefinestyle{python}{
    language=Python,
    morekeywords=[3]{%
        num_qubits,%
        duration,%
        error%
    }
}
\lstdefinelanguage{json}{
    keywords=[2]{true,false,null},
    string=[b]",
    comment=[l]{\#},
}
\lstdefinestyle{json}{
    language=json,
    stringstyle=\color{TUM_blue}
}
\lstdefinestyle{qdmi-c}{
    language=C,
    morekeywords=[2]{%
        QDMI_Session,QDMI_Job,QDMI_Device,%
        QDMI_Site,QDMI_Operation,%
        QDMI_Session_Parameter,QDMI_Job_Parameter,%
        QDMI_Device_Property,QDMI_Site_Property,%
        QDMI_Operation_Property%
    },
    morekeywords=[3]{%
        QDMI_SESSION_PARAMETER_TOKEN,%
        QDMI_SESSION_PROPERTY_DEVICES,%
        QDMI_DEVICE_PROPERTY_QUBITSNUM,%
        QDMI_JOB_PARAMETER_PROGRAM%
    },
}
\begin{document}

    \title{Practical HPCQC Integration with QDMI: \\ A Real-Hardware Case Study with IQM Systems}

    \ifshowauthors
    \author{
        \IEEEauthorblockN{
            Lukas Burgholzer\IEEEauthorrefmark{1}\IEEEauthorrefmark{2},
            Marcel Walter\IEEEauthorrefmark{1}\IEEEauthorrefmark{2},
            Patrick Hopf\IEEEauthorrefmark{1}\IEEEauthorrefmark{2},\\
            Álvaro Caride-Tabarés Sánchez\IEEEauthorrefmark{3},
            Teemu Mattsson\IEEEauthorrefmark{3},
            Bernd Hoffmann\IEEEauthorrefmark{4},
            Noora Färkkilä\IEEEauthorrefmark{3},
            Daniel Bulmash\IEEEauthorrefmark{4},\\
            Robert Wille\IEEEauthorrefmark{1}\IEEEauthorrefmark{2},
            Eric Mansfield\IEEEauthorrefmark{4}
        }
        \IEEEauthorblockA{\IEEEauthorrefmark{1}%
        MQSC,
            Garching near Munich, Germany
        }
        \IEEEauthorblockA{\IEEEauthorrefmark{2}%
        Chair for Design Automation,
            Technical University of Munich,
            Munich, Germany
        }
        \IEEEauthorblockA{\IEEEauthorrefmark{3}%
        IQM Quantum Computers,
            Espoo, Finland
        }
        \IEEEauthorblockA{\IEEEauthorrefmark{4}%
        IQM Quantum Computers,
            Munich, Germany
        }
        \{%
        lukas, %
        marcel, %
        patrick, %
        robert\}@mq.sc\\
        \{%
        alvaro.caride, %
        teemu.mattsson, %
        bernd.hoffmann, %
        noora.farkkila, %
        daniel.bulmash, %
        eric.mansfield\}@iqm.tech
    }
    \hypersetup{ %
        pdftitle={Practical HPCQC Integration with QDMI:\ A Real-Hardware Case Study with IQM Systems},
        pdfsubject={The International Conference for High Performance Computing, Networking, Storage, and Analysis 2026},
        pdfauthor={
            Lukas Burgholzer
            Marcel Walter,
            Patrick Hopf,
            Álvaro Caride-Tabarés Sánchez,
            Teemu Mattsson,
            Bernd Hoffmann,
            Noora Färkkilä,
            Daniel Bulmash,
            Robert Wille,
            Eric Mansfield
        }
    }
    \else
    \author{%
        \vspace{9em}
    }
    \hypersetup{ %
        pdftitle={},
        pdfsubject={},
        pdfauthor={}
    }
    \fi

    \maketitle
    \begin{abstract}
        Quantum computers are moving into HPC centers, and the main challenge is now integration rather than pure hardware access. Many current software paths still depend on vendor-specific adapter chains between user SDKs, schedulers, and backend APIs. This pattern makes operations more complex than necessary and slows the transition from pilots to production workflows.
        We present a practical integration path centered on the Quantum Device Management Interface (QDMI). Using IQM superconducting systems as a hardware case study, we implement an IQM-backed QDMI layer and connect it to two software layers that HPC centers working with quantum computers already care about: Slurm-based job execution and Qiskit-facing user workflows.
        The implementation is publicly available at \href{https://github.com/iqm-finland/QDMI-on-IQM}{github.com/iqm-finland/QDMI-on-IQM}.
        The key message is simple: integrating quantum hardware into HPC does not have to be a bespoke engineering effort for each backend. Once the software-hardware boundary is standardized, large parts of the stack become reusable across providers and deployment styles. Our results do not claim that standardization eliminates all HPCQC challenges. They show that this specific boundary can already be standardized today in a way that is practical for users, operators, and vendors.
    \end{abstract}


    \section{Introduction}\label{sec:intro}
    HPC centers are used to integrating new accelerators without having to rewrite their entire software stack every few years.
    The reason this works is that interfaces are stable: users keep familiar tools, operators keep manageable workflows, and hardware diversity is handled with clear system boundaries~\cite{Dagum1998,Beckingsale2019,yoo2003}.

    Quantum computing is now entering the same environment, but many integration paths remain fragile, as substantial parts of the quantum software ecosystem are not designed for HPC requirements~\cite{Beck2024,Elsharkawy2025}. Teams often build bespoke backend-specific connections from frontend SDKs to vendor APIs and then add custom scheduler glue on top. This is workable for a short proof of concept, but it becomes expensive when a center wants to support multiple vendor backends, maintain workflows over time, or share tools across centers~\cite{Weder2020,Kaya2024,McCaskey2020}.

    For HPC operators, the problem shows up quickly. Each backend-specific path brings its own authentication behavior, status model, and error semantics. Each of these differences must then be reflected in scripts, monitoring, and user support. For users, this often means that changing the backend requires changing the entire codebase and job setup, even when the scientific workflow remains unchanged.

    In practice, this is where many HPCQC efforts slow down. The challenge is not only qubit quality or algorithm maturity; it is the amount of bespoke integration code that accumulates in the middle of the stack~\cite{Schulz2022}. Recent deployments and architectural work make this explicit: operations related to calibration updates, maintenance, and scheduling semantics are now first-class concerns~\cite{Mansfield2025,nvqlink2025,ibm_hpcqc_ra2026}. At the same time, research on resource management is converging on the idea that quantum resources should be managed through familiar HPC mechanisms rather than ad hoc side channels~\cite{ibm_pasqal_qrmi2025,openqse2025}.

    This paper addresses one concrete boundary in that larger picture: the device-management interface between HPC software layers and quantum backends. We use QDMI~\cite{qdmi2024qce,burgholzer2026mqss} as the standardization target for this boundary and evaluate it through a real integration effort with IQM hardware. The goal is deliberately practical. We are not proposing a full HPCQC architecture or middleware from scratch. Instead, we show that this central boundary can be standardized in a way that immediately reduces integration effort and directly benefits other groups seeking to integrate quantum hardware into HPC environments~\cite{hopf2026}.

    The core contribution is a full implementation and integration path publicly available at \href{https://github.com/iqm-finland/QDMI-on-IQM}{github.com/iqm-finland/QDMI-on-IQM}.
    First, we build and analyze an IQM-backed QDMI implementation that maps real backend behavior into QDMI session, query, and job semantics. Second, we connect that layer to common HPC software paths: Slurm-based execution and a Qiskit-facing frontend adapter that targets QDMI rather than a single vendor API. Third, we validate the path with an end-to-end \emph{Quantum Selected Configuration Interaction}~(QSCI) workflow that runs through the same components users and operators would use in practice.

    Our central claim is intentionally modest and actionable: integrating quantum hardware into HPC need not be scary or vendor-locked. The software layers above the device-management boundary can be shared once that boundary is sufficiently standardized. IQM is the concrete case-study platform, but the integration pattern is broader than a single vendor.

    This claim does not imply that all integration work disappears. Physical devices still have vendor-specific payload formats, calibration behavior, and operational constraints. The point is that these differences can be isolated within a device plugin, rather than leaking through every layer of the HPCQC software stack.

    The remainder of the paper is structured as follows. \cref{sec:prelim} introduces the required background and operational context. \cref{sec:impl} explains the IQM QDMI implementation in detail. \cref{sec:hpc-components} describes how the standardized device layer is connected to HPCQC workflows.
    \cref{sec:deployment} elaborates three potential deployment scenarios. \cref{sec:demo} walks through an end-to-end execution of the QSCI algorithm through the proposed infrastructure. Finally, \cref{sec:conclusion} summarizes practical takeaways for centers and vendors.

    \section{Preliminaries}\label{sec:prelim}
    To keep this paper self-contained, this section reviews two central components: QDMI, the interface proposed for standardization, and IQM systems, the hardware platform targeted in our case study.

    \begin{figure}[t]
        \centering
        \includegraphics[width=\columnwidth]{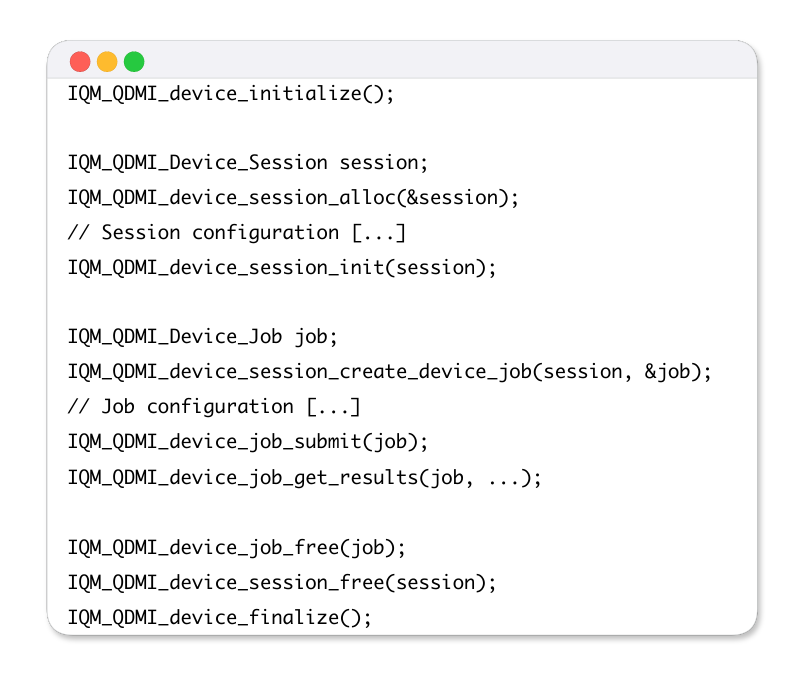}\vspace*{-4mm}
        \caption{Minimal QDMI interaction pattern; device, session, and job lifecycle.}
        \label{fig:qdmi-minimal-flow}
    \end{figure}

    \subsection{QDMI as an HPC-Facing Boundary}\label{sec:prelim:qdmi}
    QDMI~\cite{qdmi2024qce,burgholzer2026mqss} is designed to standardize the software-hardware boundary for quantum devices, providing a stable, vendor-agnostic API for backend integration.
    This device-level abstraction complements middleware solutions operating above QDMI, enabling resource scheduling, job orchestration, and unified management of quantum and classical resources in HPC environments.
    By focusing on the device interface, QDMI enables researchers to build portable, extensible solutions for hybrid quantum-classical workflows.
    Indeed, QDMI is an open standard that already supports multiple physical backends beyond the superconducting case study presented here, including trapped-ion systems (AQT) and cloud providers (Amazon Braket)~\cite{hopf2026}.

    QDMI is written in C, which makes it straightforward to wrap and implement in a large variety of languages (C++, Fortran, Rust, or Python), stable enough for long-lived integration code, and explicit about resource ownership and error handling.

    Conceptually, QDMI provides three function groups. \emph{Session} functions create and initialize an authenticated execution context for a selected backend. \emph{Query} functions expose device properties ranging from static architecture data to calibration-sensitive values. \emph{Job} functions manage program submission, job status tracking, cancellation, and result retrieval.

        \begin{figure}[t]
        \centering
        \includegraphics[width=0.75\columnwidth]{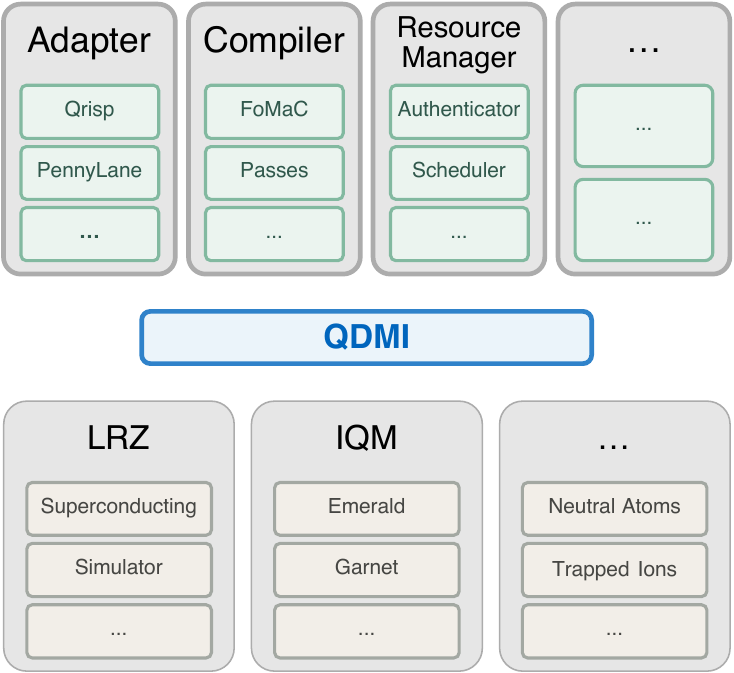}
        \caption{QDMI as the standardized software-hardware boundary, facilitating interaction between various software components and diverse hardware platforms.}
        \label{fig:qdmi-hourglass}
    \end{figure}

    \begin{example}
        \autoref{fig:qdmi-minimal-flow} demonstrates a typical QDMI interaction pattern involving the aforementioned function groups.
        Note that in addition to the leading \texttt{IQM\_} identifier, the same pattern applies to any QDMI device, regardless of the backend details.
        This figure shows the concrete symbols exported by the IQM device; the remaining figures and tables use the corresponding unprefixed QDMI interface names, as they describe the standardized boundary rather than any particular implementation.
        First, a device is initialized.
        Next, an authenticated session can be configured and established with the device.
        This session can then be used to create a job.
        After configuration, the job can be submitted for execution and its results retrieved.
        Finally, once they are no longer needed, the job and the session are explicitly freed, and the device is finalized.
    \end{example}

    The hourglass view in \autoref{fig:qdmi-hourglass} illustrates the central argument of this paper.
    The upper layers of the HPCQC software stack, such as frontend adapters, compilers, or resource managers, can evolve independently of vendor APIs as long as they target QDMI. The lower layers remain free to expose hardware-specific capabilities behind the interface.

    It is also important to be explicit about scope.
    QDMI standardizes the control boundary, not the complete scientific workflow.
    It does not decide which transpiler (or compiler) to use~\cite{burgholzer2026mqtcompiler}, which algorithm framework to pick, or how a center schedules mixed classical and quantum resources globally.
    Those concerns remain in higher layers, but they arguably become easier to build and maintain when the hardware boundary is stable.

    \subsection{IQM Systems and Server API Context}
    Our case study uses a superconducting quantum computer because such systems represent a realistic HPC integration target: operational quantum hardware, recurring calibration updates, and both remote and on-premises deployment modes~\cite{abdurakhimov2024,Rnkk2024}.
    This lets us evaluate whether QDMI is expressive enough to cover different deployment styles and whether the same software path can survive across them.
    As an established vendor in the HPCQC landscape, IQM systems come with an existing software stack into which QDMI must be embedded.
    \autoref{fig:iqm-sw-stack} shows an overview of the software stack, with QDMI serving as a shared interface between middleware components and the backend IQM Server API.
    Specifically, the left side of the diagram illustrates the client-side environment, where high-level workflows are created using front-end SDKs (such as Qiskit) and compiled or optimized with tools (such as MQT Core) into a backend-compliant format, before being passed through QDMI bindings to the device layer.

    \begin{figure}[t]
        \centering
        \resizebox{\columnwidth}{!}{
            \definecolor{SWpurple}{RGB}{72,65,168}
\definecolor{SWpurplefill}{RGB}{233,231,248}
\definecolor{SWgreen}{RGB}{68,144,108}
\definecolor{SWgreenfill}{RGB}{225,242,234}
\definecolor{SWblue}{RGB}{63,94,181}
\definecolor{SWbluefill}{RGB}{224,236,252}
\definecolor{SWgrey}{RGB}{140,148,144}
\definecolor{SWgreyfill}{RGB}{199,207,204}
\definecolor{SWpanel}{RGB}{233,235,234}
\definecolor{SWbeige}{RGB}{241,239,230}
\definecolor{SWink}{RGB}{53,53,51}

\begin{tikzpicture}[
    x=1.0cm, y=1.0cm,
    font=\sffamily\tiny,
    text=SWink,
    >=Stealth,
    line width=0.4pt,
    panel/.style={
        draw=SWgreen!70, densely dashed, line width=0.5pt,
        rounded corners=3pt, fill=SWpanel,
    },
    panelpurple/.style={panel, draw=SWpurple!55},
    subpanel/.style={
        draw=SWgreen!55, densely dashed, line width=0.5pt,
        rounded corners=3pt, fill=SWgreyfill,
    },
    subpanelpurple/.style={
        draw=SWpurple!55, densely dashed, line width=0.5pt,
        rounded corners=3pt, fill=SWpurplefill,
    },
    ptitle/.style={font=\sffamily\tiny\bfseries, anchor=west,
                   inner xsep=1.2pt, inner ysep=0.6pt},
    ctitle/.style={ptitle, anchor=center},
    box/.style={draw, rounded corners=2pt, align=center,
                inner xsep=2pt, inner ysep=1.6pt},
    sdk/.style={box, draw=SWgreen!70, fill=SWgreenfill},
    connblue/.style={box, draw=SWblue!70, fill=SWbluefill, text=SWblue!85!black},
    connpurple/.style={box, draw=SWpurple!55, fill=SWpurplefill,
                       text=SWpurple!80!black},
    fmt/.style={box, draw=SWgrey, fill=SWgreyfill, font=\sffamily\tiny\bfseries},
    fmtpurple/.style={box, draw=SWpurple!55, fill=SWpurplefill,
                      font=\sffamily\tiny\bfseries, text=SWpurple!80!black},
    qdmi/.style={box, draw=SWgrey, fill=SWbeige},
    server/.style={box, draw=SWgreen!80, fill=SWgreenfill, rounded corners=0.3cm},
    badge/.style={
        draw=none, fill=SWgreen!88!black, text=white, rounded corners=1.4pt,
        inner xsep=1.4pt, inner ysep=0.7pt, font=\sffamily\tiny\bfseries,
        scale=0.52, transform shape, anchor=north east,
    },
    flow/.style={draw=SWgreen, ->, line width=0.4pt},
    trunk/.style={draw=SWgreen, line width=0.4pt},
    flowd/.style={draw=SWgreen, ->, line width=0.4pt, densely dashed},
    trunkd/.style={draw=SWgreen, line width=0.4pt, densely dashed},
    flowp/.style={draw=SWpurple, ->, line width=0.4pt},
    sepline/.style={draw=SWgrey!55, dotted, line width=0.35pt},
]

\def\lnA{-2.28}   %
\def\lnB{-2.50}   %
\def\lnC{-2.72}   %
\def\lnD{-2.94}   %
\def\lnE{-3.16}   %
\def\lnF{-3.38}   %

\draw[panelpurple] (0.15,-0.15) rectangle (2.30,-2.05);
\draw[panel]       (2.50,-0.15) rectangle (10.55,-2.05);
\draw[panel]       (0.15,-3.62) rectangle (10.55,-9.15);
\draw[panel]       (0.15,-9.40) rectangle (10.55,-10.52);

\draw[sepline] (3.75,-3.70) -- (3.75,-9.07);
\draw[sepline] (6.90,-3.70) -- (6.90,-9.07);

\draw[subpanel]       (7.25,-4.42) rectangle (10.45,-5.42);
\draw[subpanelpurple] (7.30,-5.70) rectangle (10.40,-6.80);
\draw[subpanel]       (4.15,-7.95) rectangle (9.75,-8.95);

\node[sdk, minimum width=1.60cm]        (experiments) at (1.20,-0.90) {Experiments};
\node[connpurple, minimum width=1.60cm] (exacore)     at (1.20,-1.62) {EXA Core};

\node[sdk, minimum width=1.10cm] (pennylane) at (4.20,-0.90) {Pennylane};
\node[sdk, minimum width=0.95cm] (qrisp)     at (5.45,-0.90) {QRISP};
\node[sdk, minimum width=1.05cm] (cudaq)     at (7.10,-0.90) {CUDA-Q};

\node[sdk, minimum width=0.80cm] (cirq)   at (3.30,-1.62) {Cirq};
\node[sdk, minimum width=0.90cm] (qiskit) at (4.55,-1.62) {Qiskit};
\node[sdk, minimum width=1.45cm] (pulla)  at (6.05,-1.62) {IQM Pulla};

\node[box, draw=SWgreen!70, fill=SWgreenfill, minimum width=2.70cm]
     (iqmclient) at (2.35,-4.80) {\textbf{IQM Client}\\Qiskit on IQM\\Cirq on IQM};
\node[connblue, minimum width=2.20cm]
     (mqtcore) at (5.05,-4.80) {\textbf{MQT Core}\\Munich Quantum Toolkit\\circuit compilation};

\node[sdk, minimum width=1.10cm] (adaqsk) at (7.90,-5.02) {Qskt/CQ\\adapters};
\node[sdk, minimum width=0.95cm] (adapny) at (9.85,-5.02) {Pennylane\\adapter};

\node[connpurple, minimum width=2.70cm] (scc) at (1.70,-5.90)
     {\textbf{Station Control Client}};
\node[connpurple, minimum width=1.60cm] (mqsspasses) at (8.75,-6.30)
     {\textbf{MQSS Passes}\\QIR / Quake $\cdot$ LLVM};

\node[fmtpurple, minimum width=1.05cm] (fmtpulse)  at (1.00,-7.25) {IQM Pulse\\Playlist};
\node[fmt, minimum width=1.70cm]       (fmtclient) at (2.75,-7.25) {IQM JSON / QIR\\IQM Pulse Playlist};
\node[fmt, minimum width=1.05cm]       (fmtjson)   at (5.05,-7.25) {IQM JSON};
\node[fmt, minimum width=0.55cm]       (fmtqir)    at (8.75,-7.25) {QIR};

\node[qdmi, minimum width=5.20cm] (qdmiiqm) at (6.95,-8.55)
     {\textbf{QDMI on IQM}\\IQM's QDMI device implementation};

\node[server, minimum width=2.20cm, minimum height=0.55cm] (iqmserver)
     at (4.30,-10.05) {\textbf{IQM Server}\\quantum hardware};

\draw[flowp] (experiments) -- (exacore);

\draw[trunkd] (pennylane.south) |- (4.55,-1.25);
\draw[trunkd] (qrisp.south)     |- (4.55,-1.25);
\draw[flowd]  (4.55,-1.25) -- (qiskit.north);

\draw[trunk] (pulla.south)  |- (1.40,\lnA);
\draw[trunk] (qiskit.south) |- (1.40,\lnB);
\draw[trunk] (cirq.south)   |- (1.40,\lnC);
\draw[flow]  (1.40,\lnA) -- ($(iqmclient.north west)+(0.40,0)$);

\draw[flow]  (qiskit.south) |- (4.80,\lnD) -| (mqtcore.north);

\draw[trunk] (cudaq.south)  |- (7.65,\lnE);
\draw[trunk] (qiskit.south) |- (7.65,\lnF);
\draw[flow]  (7.65,\lnE) -- ($(adaqsk.north west)+(0.30,0)$);

\draw[flow] (pennylane.north) |- (6.20,0.22) -| (adapny.north);

\draw[flowp] (exacore.south) -- (1.20,-1.90) -- (0.75,-1.90) -- ($(scc.north west)+(0.40,0)$);

\draw[flow]  ($(iqmclient.south west)+(2.30,0)$) -- ($(fmtclient.north west)+(1.40,0)$);
\draw[flowp] ($(scc.south west)+(0.65,0)$) -- (fmtpulse.north);
\draw[flow]  (mqtcore.south) -- (fmtjson.north);

\draw[trunk] (adaqsk.south) |- (8.75,-5.55);
\draw[trunk] (adapny.south) |- (8.75,-5.55);
\draw[flow]  (8.75,-5.55) -- (mqsspasses.north);
\draw[flow]  (mqsspasses.south) -- (fmtqir.north);

\draw[flow] (fmtjson.south) -- ($(qdmiiqm.north west)+(0.70,0)$);
\draw[flow] (fmtqir.south)  -- ($(qdmiiqm.north west)+(4.40,0)$);

\draw[flow]  (qdmiiqm.south) -- (6.95,-9.96) -- ($(iqmserver.east)+(0,0.09)$);
\draw[flow]  ($(fmtclient.south west)+(0.50,0)$) -- (2.40,-9.96) -- ($(iqmserver.west)+(0,0.09)$);
\draw[flowp] (fmtpulse.south) -- (1.00,-10.14) -- ($(iqmserver.west)+(0,-0.09)$);

\draw[flowd] (cudaq.north) |- (7.60,0.40) -| (11.45,-10.14)
             -- ($(iqmserver.east)+(0,-0.09)$);

\node[fmt, minimum width=1.05cm] (fmtdirect) at (11.45,-7.25)
     {IQM JSON\\\textit{Direct}};

\node[ptitle] at (0.28,-0.34) {Experiments};
\node[ptitle] at (2.63,-0.34) {Level 1 $\cdot$ SDKs};

\node[ptitle, fill=SWpanel] at (0.28,-3.82) {Level 2 $\cdot$ Connectors};
\node[ctitle] at (3.00,-4.14) {IQM Client flow};              %
\node[ctitle] at (6.10,-4.14) {MQT Core flow};                %
\node[ctitle] at (8.75,-4.14) {MQSS / QDMI flow};             %

\node[ctitle] at (8.75,-4.57) {Adapters};                     %
\node[ctitle] at (7.78,-5.85) {Compiler};                     %
\node[ctitle] at (6.65,-8.10) {Device interface};             %
\node[ctitle] at (4.30,-9.59) {Server side software};         %

\node[badge] at ($(experiments.north east)+(-0.02,-0.02)$) {IQM};
\node[badge] at ($(exacore.north east)+(-0.02,-0.02)$)     {IQM};
\node[badge] at ($(pulla.north east)+(-0.02,-0.02)$)       {IQM};
\node[badge] at ($(iqmclient.north east)+(-0.02,-0.02)$)   {IQM};
\node[badge] at ($(scc.north east)+(-0.02,-0.02)$)         {IQM};
\node[badge] at ($(qdmiiqm.north east)+(-0.02,-0.02)$)     {IQM};

\node[badge, anchor=west] at (10.25,-9.5) {IQM};

\end{tikzpicture}
        }
        \caption{A high-level architecture diagram showing three abstraction layers for the IQM software stack. Quantum circuits are submitted from level 1 as Qiskit~\cite{qiskit} code to MQT Core~\cite{burgholzer2025MQTCore, mqt}, which converts the circuits to IQM JSON and sends them to QDMI-on-IQM. This interfaces with the IQM Server API.}
        \label{fig:iqm-sw-stack}
    \end{figure}

    At the lowest level sits the IQM Server API, which exposes the backend primitives needed for the QDMI device implementation.
    It allows discovering a list of available backends that can be exposed through QDMI.
    On the query side, it exposes static architecture information (\eg, device topology, site identifiers, and operational availability) and dynamic quality metrics (\eg, qubit coherence times and gate fidelities).
    On the execution side, it provides job submission, status inspection, artifact access, and, where applicable, cancellation behavior.
    Every IQM backend fronts the hardware with its own built-in FIFO queue, so submissions from concurrent clients are accepted and executed in order rather than contending for the device.
    Under a scheduler-managed deployment, a job therefore traverses two queues in sequence, the center's Slurm queue and then the device's own queue, which is the operational fact behind the queue-length and queue-position properties discussed in \autoref{sec:impl} and behind the optional Slurm-license mechanism of \autoref{sec:hpc-components:scheduler}.

    In the following section, we will describe how these concepts map to QDMI and how they can be successfully abstracted behind this common interface.

    \begin{table*}[!t]
        \caption{Functional mapping between QDMI entry points and IQM Server API.}
        \label{tab:functional-groups}
        \centering
        \resizebox{\textwidth}{!}{%
        \renewcommand{\arraystretch}{1.5}
        \setlength{\tabcolsep}{4pt}
        \rowcolors{2}{lightgray!15}{white}
        \begin{tabular}{
            >{\raggedright\arraybackslash}p{0.42\linewidth}
            >{\raggedright\arraybackslash}p{0.58\linewidth}
        }
    \toprule
    \textbf{QDMI Function(s)} & \textbf{IQM Server API Interaction} \\
    \midrule

    \texttt{QDMI\_device\_session\_init()} &
    \textbf{Discovery and session warm-up.}
    Performs remote backend discovery, resolves the selected target by ID, alias, or first available backend, materializes static architecture, loads dynamic architecture for the default calibration set, loads calibration metrics, and probes whether calibration jobs are supported. \\

    \texttt{QDMI\_device\_session\_query\_device\_property()}, \newline
    \texttt{QDMI\_device\_session\_query\_site\_property()}, \newline
    \texttt{QDMI\_device\_session\_query\_operation\_property()} &
    \textbf{Cache-backed metadata access.}
    For architecture and calibration metadata, these functions return the already-normalized, cached session state without triggering remote API calls or refreshes. The sole exception is the device queue length, which is resolved by a live queue-availability request on every query, since a cached queue depth would be of no use to a scheduler. \\ \midrule

    \texttt{QDMI\_device\_session\_\newline retrieve\_device\_job\_by\_id()} &
    \textbf{Job reattachment.}
    Resolves an existing IQM circuit job into a QDMI job handle by reading its status endpoint, so a later process or scheduler step can attach to an earlier-submitted job without holding the original handle. The retrieved handle supports access to status, queue position, and result, but reports \texttt{QDMI\_ERROR\_NOTSUPPORTED} for parameter-derived properties it never held. \\ \midrule

    \texttt{QDMI\_device\_job\_submit()} &
    \textbf{Job submission.}
    Builds the IQM payload from QDMI job state and submits it through the standard circuit-execution path or, for calibration jobs, through the calibration-specific service path. Calibration submission is allowed only when support is detected during session initialization. \\

    \texttt{QDMI\_device\_job\_check()}, \newline
    \texttt{QDMI\_device\_job\_wait()}, \newline
    \texttt{QDMI\_device\_job\_cancel()}, \newline
    \texttt{QDMI\_device\_job\_query\_property()} &
    \textbf{Job lifecycle management.}
    For submitted jobs, \texttt{check()} polls the circuit-job or calibration-job status endpoint, depending on the job type, and maps native IQM states to QDMI job semantics. \texttt{wait()} repeatedly invokes \texttt{check()} with exponential backoff, while \texttt{cancel()} uses the corresponding circuit-cancel or calibration-abort path and updates local job state accordingly. Querying the queue position follows the same path: it invokes \texttt{check()} to refresh the status and position, and returns a value only while the job is queued. \\

    \texttt{QDMI\_device\_job\_get\_results()} with \texttt{QDMI\_JOB\_RESULT\_SHOTS}, \newline
    \texttt{QDMI\_JOB\_RESULT\_HIST\_KEYS}, \newline
    \texttt{QDMI\_JOB\_RESULT\_HIST\_VALUES}, or \newline
    \texttt{QDMI\_JOB\_RESULT\_CUSTOM1} &
    \textbf{Job-result retrieval.}
    For completed circuit jobs, the first uncached access fetches measurement artifacts or measurement counts, derives histogram data locally when non-empty shot data is already cached, converts the result into QDMI formats, and caches it for subsequent accesses.\newline
    For completed calibration jobs with \texttt{QDMI\_JOB\_RESULT\_CUSTOM1}, the first uncached access rereads calibration-job status, extracts the new calibration-set identifier, refreshes the session's dynamic architecture and calibration metrics to that calibration set, and retries that refresh once after 120\,s if the first attempt fails. \\
    \bottomrule
    \end{tabular}
    }
    \end{table*}

    \section{Implementing the IQM QDMI Device}\label{sec:impl}
    With the high-level context in place, this section describes the implementation in the IQM-backend QDMI device and explains the rationale for specific engineering choices. The goal is to make the integration path reproducible, not to present a conceptual sketch. The implementation is available as open source at \href{https://github.com/iqm-finland/QDMI-on-IQM}{github.com/iqm-finland/QDMI-on-IQM}, enabling other centers to inspect and adapt the pattern rather than rebuilding it from scratch.
    \autoref{tab:functional-groups} summarizes the functional implementation presented in this section.

    \begin{figure}[t]
        \centering
        \resizebox{\columnwidth}{!}{
            \begin{tikzpicture}[
    font=\sffamily\footnotesize,
    node distance=0.35cm,
    >=Stealth,
    qdmi boundary/.style={
        draw=TUM_blue, line width=1.2pt,
        fill=TUM_blue!8,
        rounded corners=3pt,
        minimum width=7.6cm, minimum height=0.7cm,
        font=\sffamily\footnotesize\bfseries,
        text=TUM_blue
    },
    qdmi func/.style={
        draw=IQM_darkgreen!70, line width=0.6pt,
        fill=IQM_darkgreen!12,
        rounded corners=2pt,
        minimum width=2.1cm, minimum height=0.65cm,
        font=\sffamily\scriptsize,
        text=IQM_verydarkgreen,
        align=center
    },
    support layer/.style={
        draw=IQM_brown, line width=0.6pt,
        fill=IQM_lightbrown,
        rounded corners=2pt,
        minimum width=2.1cm, minimum height=0.65cm,
        font=\sffamily\scriptsize,
        text=IQM_verydarkbrown,
        align=center
    },
    group frame/.style={
        draw=TUM_darkblue!30, line width=0.7pt,
        densely dashed, rounded corners=4pt,
        inner sep=5pt,
    },
    support frame/.style={
        draw=IQM_brown, line width=0.7pt,
        densely dashed, rounded corners=4pt,
        inner sep=5pt,
    },
    vendor api/.style={
        draw=TUM_grey!80, line width=1.2pt,
        fill=TUM_lightgrey!50,
        rounded corners=3pt,
        minimum width=7.6cm, minimum height=0.7cm,
        font=\sffamily\footnotesize\bfseries,
        text=TUM_black!80
    },
    endpoint/.style={
        draw=TUM_grey!50, line width=0.4pt,
        fill=TUM_white,
        rounded corners=1.5pt,
        minimum width=2.2cm, minimum height=0.55cm,
        font=\sffamily\tiny,
        text=TUM_black!70,
        align=center
    },
    arrow thick/.style={
        -{Stealth[length=5pt, width=3.5pt]},
        line width=0.8pt, color=IQM_darkgreen!60
    },
    arrow grey/.style={
        -{Stealth[length=4pt, width=3pt]},
        line width=0.7pt, color=TUM_grey!60
    },
    boundary line/.style={
        line width=1.0pt, densely dashed, color=TUM_blue!50
    },
]

\node[qdmi boundary] (qdmi-if) {QDMI Device Interface};

\node[qdmi func, below=0.8cm of qdmi-if.south west, anchor=north west, xshift=0.45cm]
    (session) {Session\\[-1pt]{\tiny init $\cdot$ alloc $\cdot$ free}};
\node[qdmi func, right=0.2cm of session]
    (query) {Query\\[-1pt]{\tiny device $\cdot$ site $\cdot$ op}};
\node[qdmi func, right=0.2cm of query]
    (job) {Job\\[-1pt]{\tiny submit $\cdot$ check $\cdot$ results}};

\node[group frame, fit=(session)(query)(job), label={[font=\sffamily\tiny\itshape, text=IQM_darkgreen!60]above left:QDMI functions}]
    (func-group) {};

\node[support layer, below=1.1cm of session]
    (token) {Token\\Manager};
\node[support layer, right=0.2cm of token]
    (apiconf) {API Config\\Module};
\node[support layer, right=0.2cm of apiconf]
    (http) {HTTP\\Transport};

\node[support frame, fit=(token)(apiconf)(http), label={[font=\sffamily\tiny\itshape, text=IQM_brown]above left:Support layer}]
    (support-group) {};

\draw[arrow thick, line width=1.0pt] (func-group.south) -- (support-group.north);

\node[vendor api, below=0.8cm of support-group.south, anchor=north]
    (iqm-api) {IQM Server API \textnormal{\scriptsize(REST/HTTPS)}};

\draw[arrow thick] (support-group.south) -- (iqm-api.north);

\node[endpoint, anchor=north west] at ([yshift=-0.30cm, xshift=0.15cm]iqm-api.south west)
    (ep-disc) {Device\\[-1pt]Discovery};
\node[endpoint, anchor=north] at ([yshift=-0.30cm]iqm-api.south)
    (ep-arch) {Architecture \& \\[-1pt] Calibration};
\node[endpoint, anchor=north east] at ([yshift=-0.30cm, xshift=-0.15cm]iqm-api.south east)
    (ep-jobs) {Job\\[-1pt]Lifecycle};

\draw[arrow grey] ([xshift=-2.4cm]iqm-api.south) -- (ep-disc.north);
\draw[arrow grey] (iqm-api.south) -- (ep-arch.north);
\draw[arrow grey] ([xshift=2.4cm]iqm-api.south) -- (ep-jobs.north);

\begin{scope}[on background layer]
    \node[fill=IQM_lightgreen!50, rounded corners=5pt, inner sep=10pt,
          fit=(func-group)(support-group)]
        (plugin-bg) {};
\end{scope}
\node[font=\sffamily\tiny, text=IQM_darkgreen!80, anchor=north west]
    at (plugin-bg.south west) {IQM QDMI Device};

\end{tikzpicture}
        }
        \caption{Architecture of the IQM QDMI device implementation.}
        \label{fig:iqm-device-arch}
    \end{figure}

    \subsection{Implementation Architecture}
    The IQM QDMI device is built as a shared library that exposes all QDMI functions for session management, property queries, and job handling.
    In the implementation, most of the device logic is concentrated in a single core implementation file, while authentication, endpoint construction, and HTTP transport are factored into a small support layer. \autoref{fig:iqm-device-arch} summarizes this structure.

    The implementation focuses on two internal data structures: the session object and the job object.
    The session stores connection-related information, including authentication parameters, the HTTP client and its connection pool, the token manager, and the API configuration.
    It also holds the selected quantum computer identifier, the active calibration set identifier, and the metadata cached during session initialization.
    The job object stores a reference to its session, along with the submitted program information, backend-specific execution options, the current job status, and cached result data.
    Most of the QDMI calls operate internally on these two objects.

    The support layer is deliberately small.
    A token manager resolves credentials and supplies the bearer token used for all requests to the IQM Server API.
    The API configuration module maps symbolic backend operations, such as static-architecture queries and dynamic-architecture queries for calibration-sensitive values that may change at runtime, as well as job submission and result retrieval, to concrete IQM Server API routes.
    Finally, the HTTP transport implements a simple \texttt{GET}/\texttt{POST} interface on top of \texttt{cpr}, a C++ wrapper around \texttt{libcurl}, that performs requests and translates HTTP responses into QDMI status codes.
    Each session owns a connection pool that all its requests share, so the TCP and TLS handshake cost is paid once rather than per call.

    A key requirement for the entire architecture is that standardization must not come at the price of performance.
    In particular, it must not add unnecessary latency compared to calling the vendor-specific IQM Server API directly.
    As will become clear in the following sections, there is substantial overlap between the capabilities of the QDMI and IQM Server APIs.
    This allows the QDMI implementation to form a thin wrapper around proprietary calls without heavy processing in between, thereby meeting the requirement for negligible overhead.

    \begin{figure*}[t]
        \centering
        \includegraphics[width=\linewidth]{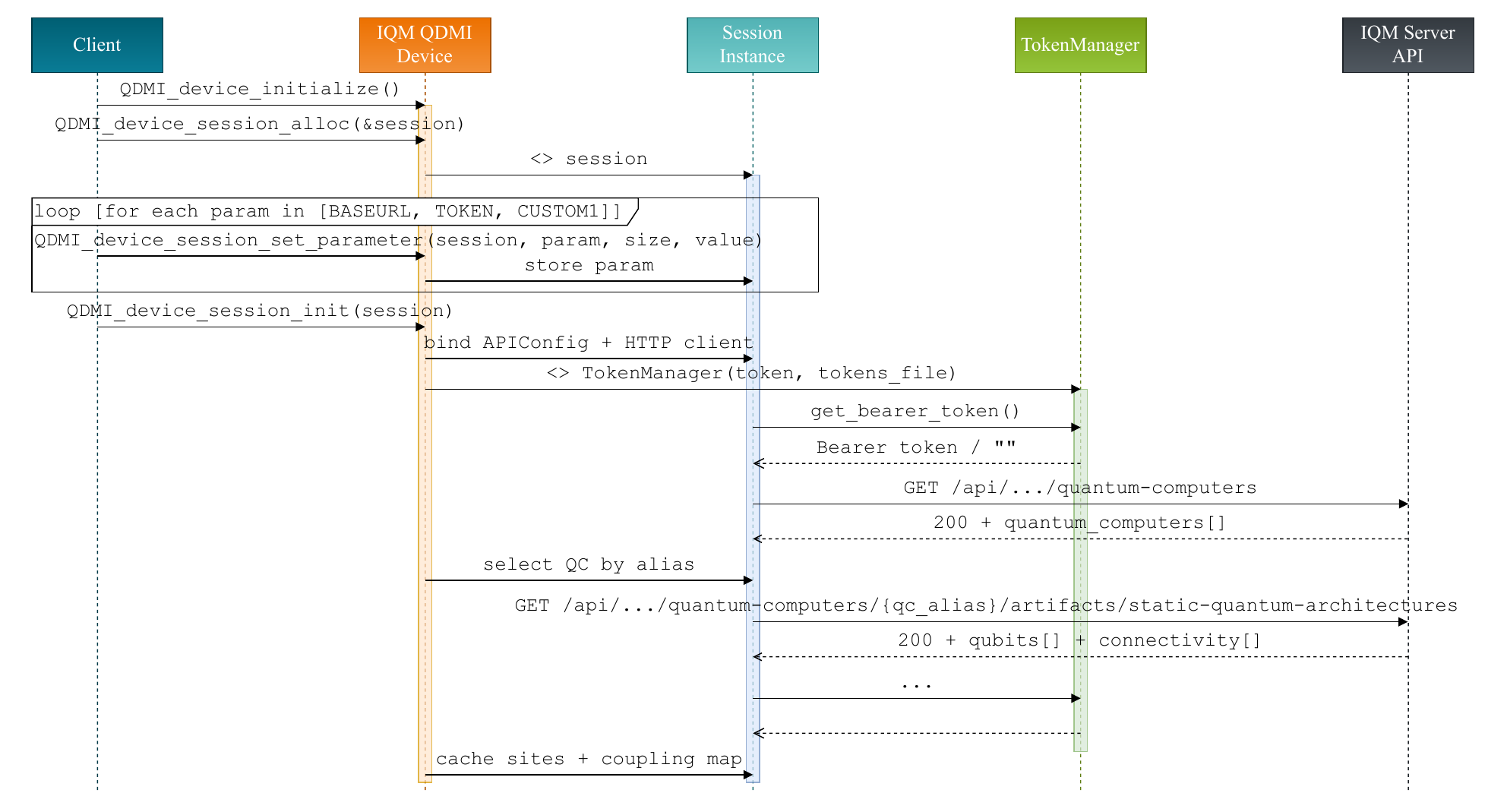}
        \caption{IQM QDMI device session setup sequence and resulting interaction with IQM Server API.}
        \label{fig:session-sequence}\vspace*{0.5em}
    \end{figure*}

    \begin{figure}[hb!]
        \centering
        \includegraphics[width=\columnwidth]{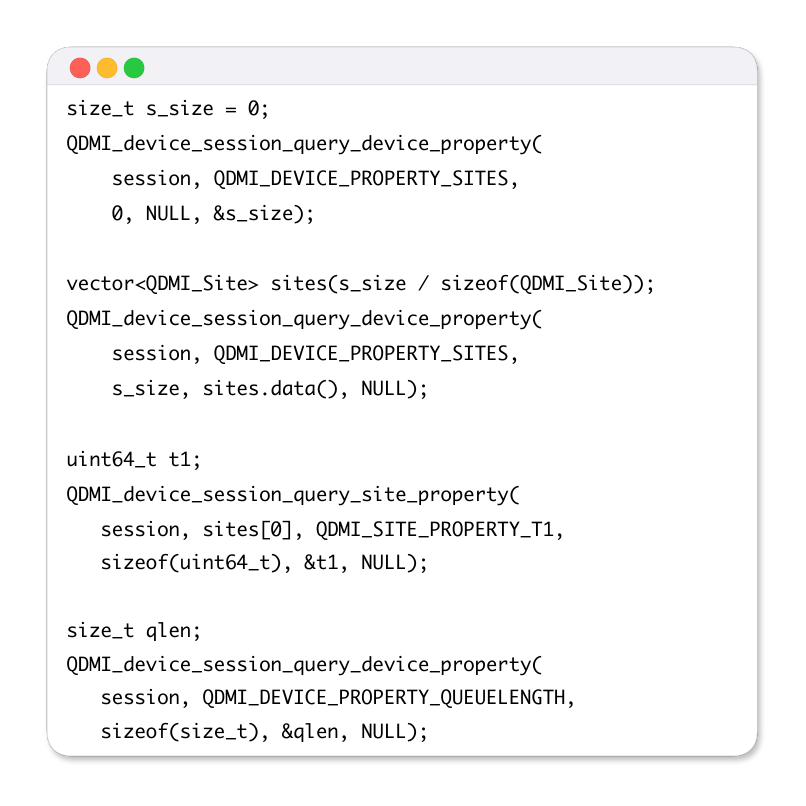}
        \caption{Property query sequence for the IQM QDMI device.}
        \label{fig:iqm-query}
    \end{figure}

    \begin{table}[!b]
        \caption{Session and job parameters and queried queue properties in the IQM QDMI device implementation.}\label{tab:parameters}
        \centering
        \scriptsize
        \renewcommand{\arraystretch}{1.1}
        \setlength{\tabcolsep}{4pt}
        \rowcolors{2}{lightgray!15}{white!15}
        \begin{tabular}{
            >{\raggedright\arraybackslash}p{0.22\linewidth}
            >{\raggedright\arraybackslash}p{0.68\linewidth}
        }
            \toprule
            \textbf{QDMI Parameter} & \textbf{Description} \\
            \midrule
            \multicolumn{2}{l}{\textit{Session Parameters}} \\
            \texttt{BASEURL}
            & Base URL of the IQM Server. This must be set before \texttt{QDMI\_device\_session\_init}. \\
            \texttt{TOKEN}
            & Explicit bearer token. \\
            \texttt{AUTHFILE}
            & Path to the IQM tokens file. \\
            \texttt{CUSTOM1}
            & Quantum computer ID selector. \\
            \texttt{CUSTOM2}
            & Quantum computer alias selector. If neither ID nor alias is set, the first available QC is used. \\
            \texttt{CUSTOM3}
            & Per-request HTTP timeout, in milliseconds. \\ \midrule
            \multicolumn{2}{l}{\textit{Job Parameters}} \\
            \texttt{PROGRAMFORMAT}
            & Selects the program format. Supported values are \texttt{IQMJSON}, \texttt{QIRBASESTRING}~\cite{stade2025qir}, and \texttt{CALIBRATION} when calibration jobs are available. \\
            \texttt{PROGRAM}
            & Program payload. This must be set before \texttt{QDMI\_device\_job\_submit}. \\
            \texttt{SHOTSNUM}
            & Number of shots. \\
            \texttt{CUSTOM1}
            & Heralding mode. Accepted values are \texttt{"none"} and \texttt{"zeros"}. \\
            \texttt{CUSTOM2}
            & Move-gate validation mode. Accepted values are \texttt{"strict"}, \texttt{"allow\_prx"}, and \texttt{"none"}. \\
            \texttt{CUSTOM3}
            & Move-gate frame-tracking mode. Accepted values are \texttt{"full"}, \texttt{"no\_detuning\_correction"}, and \texttt{"none"}. \\
            \texttt{CUSTOM4}
            & Dynamic-decoupling mode. Accepted values are \texttt{"disabled"} and \texttt{"enabled"}. \\
            \texttt{CUSTOM5}
            & Qubit mapping in the form \texttt{logical:physical,logical:physical,...}. Primarily relevant for QIR jobs. \\
            \texttt{CUSTOM5+1--\newline CUSTOM5+3}
            & Additional execution controls: maximum circuit duration relative to $T_2$ (\texttt{double}), number of active-reset cycles (\texttt{size\_t}), and dynamic-decoupling strategy (string), respectively. QDMI defines five custom job parameters; the implementation accepts three further offsets beyond \texttt{CUSTOM5} for these controls. \\ \midrule
            \multicolumn{2}{l}{\textit{Queried Queue Properties}} \\
            \texttt{QUEUELENGTH}
            & Device property. Jobs waiting in the selected backend's queue, read live from the IQM Server. Unsupported on backends without queue availability. \\
            \texttt{QUEUEPOSITION}
            & Job property. Jobs ahead of a queued job. Every query refreshes the job status, and a value is defined only while the job is queued. \\
            \bottomrule
        \end{tabular}
    \end{table}

    \subsection{Session Bootstrapping and Authentication}
\begin{figure*}[ht!]
        \centering
        \includegraphics[width=\linewidth]{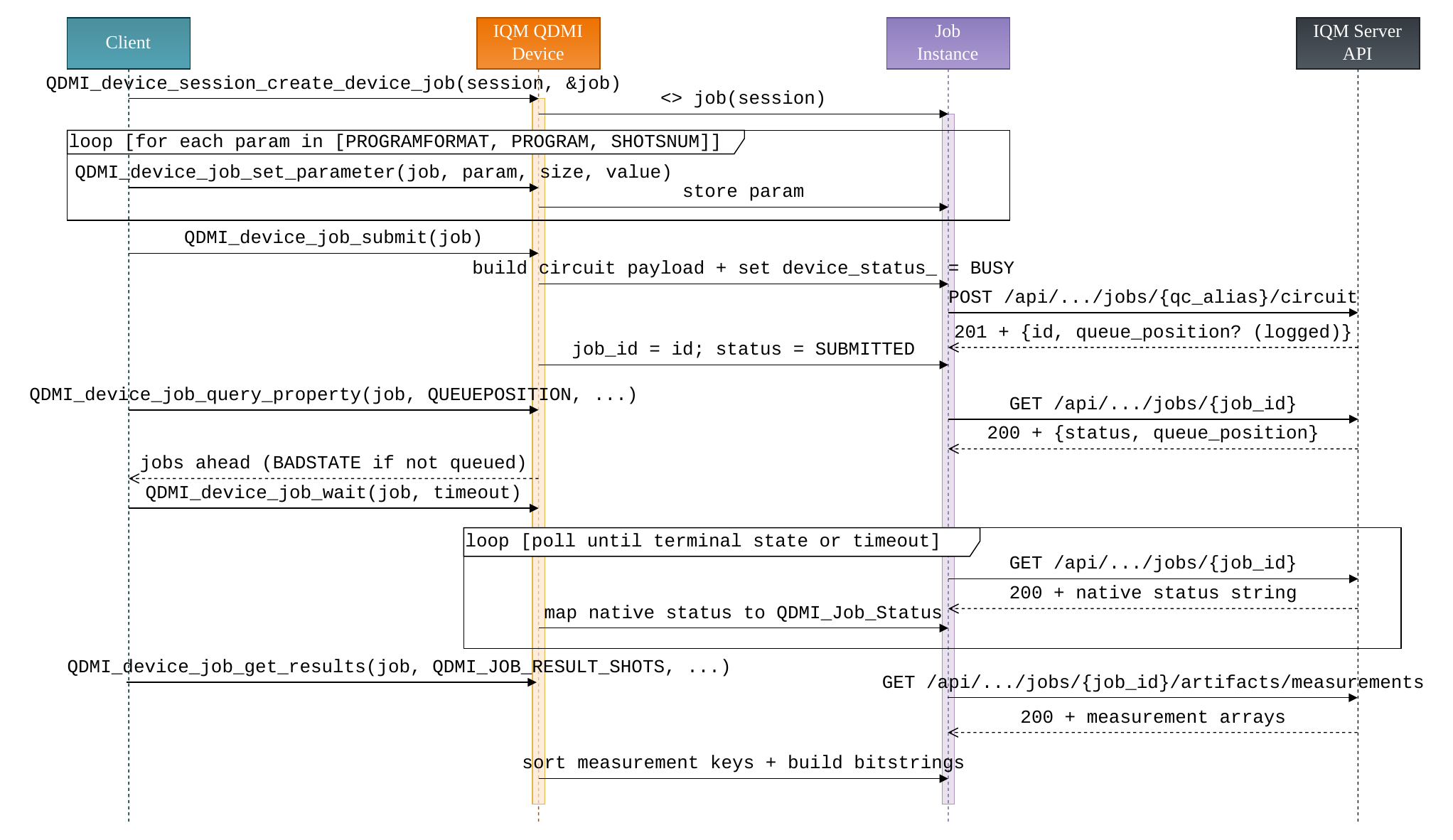}\vspace*{-3mm}
        \caption{IQM QDMI device job lifecycle sequence and resulting interaction with IQM Server API.}
        \label{fig:job-sequence}\vspace*{-3mm}
    \end{figure*}

    The first cornerstone and entry point to QDMI is the implementation of the QDMI session functions.
    In this context, a session represents an authenticated connection to the IQM Server plus a resolved target backend.
    \autoref{fig:session-sequence} shows a typical session setup workflow, using the configuration parameters listed in \autoref{tab:parameters}.\footnote{Constant names are abbreviated throughout: a session parameter written \texttt{X} denotes \texttt{QDMI\_DEVICE\_SESSION\_PARAMETER\_X}, a job parameter written \texttt{X} denotes \texttt{QDMI\_DEVICE\_JOB\_PARAMETER\_X}, a device property written \texttt{X} denotes \texttt{QDMI\_DEVICE\_PROPERTY\_X}, a job property written \texttt{X} denotes \texttt{QDMI\_DEVICE\_JOB\_PROPERTY\_X}, and a program format written \texttt{X} denotes \texttt{QDMI\_PROGRAM\_FORMAT\_X}. \autoref{tab:parameters} and \autoref{fig:session-sequence}--\ref{fig:job-sequence} use this abbreviation.}
    The session configuration accepts a base endpoint, a credential source, and an optional backend selector.
    Initializing the session then triggers API configuration, HTTP client setup, and token manager instantiation.

    First, the token manager validates the authentication configuration and supplies valid bearer tokens for all subsequent API calls to the IQM Server.
    Second, the authenticated session queries the IQM Server for a target backend
    using an explicit selector (via custom QDMI parameters, as shown in \autoref{tab:parameters}) or a default device when no selector is provided.
    Third, it retrieves initial architecture and quality metadata and stores a normalized representation in the session state.

    Since HPC systems heavily rely on injecting credentials via environment variables, authentication information can be passed either explicitly through QDMI session parameters or via environment variables.
    As a result, the same code can run unchanged, \eg, in an interactive terminal or as a Slurm batch job~\cite{yoo2003}.

    \subsection{Capability and Calibration Queries}
    The second cornerstone of the implementation and the primary mechanism through which a caller collects device, site, and operation properties is the query functions.
    Queries span three classes of data. Static architecture data covers device topology, site identifiers, and operation availability. Dynamic quality data covers calibration-sensitive metrics, such as qubit coherence times and gate fidelity indicators. Live operational state covers values that describe the backend's current load rather than its capabilities; at present, this is the device queue length.

    All three classes are normalized into QDMI properties, so the upper layers never parse vendor payloads directly. Normalization includes unit and type conversion, as needed, so QDMI callers receive stable and comparable values. This is one of the most important portability wins in practice: compilers, transpilation passes, and scheduler-side checks can share a single query format across backends.

    \begin{example}
        The procedure shown in \autoref{fig:iqm-query} illustrates the query patterns available through the same interface function.
        The first pattern is used for properties whose size is not known in advance: the caller passes a null pointer and a size of zero to obtain the required buffer size.
        A second call then uses the returned size to allocate a buffer and retrieve the actual data, in this case a vector of site (that is, qubit) identifiers.
        The third query demonstrates direct retrieval of a fixed-size property by obtaining the T1 value for a specific site in a single call.
        The final query uses the same fixed-size pattern to read the device queue length, illustrating that the call signature does not change even though this value is resolved by a backend request rather than served from the session cache.
    \end{example}

    Data freshness is handled differently for each of the three classes.
    Static data is loaded once during initialization and only reloaded when the session is reinitialized or switched to a different backend.
    Dynamic data is also cached during session initialization, but can be refreshed by initiating calibration jobs.
 Live operational state is never cached: it is resolved by a request to the IQM Server on every query, because a queue depth that was accurate at session setup carries no information for a scheduler deciding where to place work now.
    Together, this avoids stale data without requiring every query to incur an expensive network round-trip.

    This separation between cached and refreshed data can be useful in batch-oriented HPC setups, for example, when consecutive jobs are launched through a scheduler.
    Metadata that is unlikely to change during a run over multiple jobs can be cached once, while values that may drift over time can be refreshed after each execution.
    For example, if updated error rates indicate that the quality of certain qubits has degraded, subsequent jobs may need to be rerouted around these parts of the device, which can in turn increase swap overhead and thus runtime.
 Dynamic queries thereby make runtime costs more predictable and allow later jobs to react to relevant device updates.

    \subsection{Job Lifecycle, Status Mapping, and Results}
    Finally, job handling represents the third cornerstone of the implementation.
    We map proprietary IQM Server endpoints to standardized QDMI job functions.
    \autoref{fig:job-sequence} illustrates a typical job lifecycle, in which a client creates a job handle, specifies the program and execution parameters, and requests job submission via QDMI functions.
    Internally, this triggers device-status updates and authentication via the token manager (omitted from the figure for simplicity), followed by a submission call to the IQM Server API.
    Subsequent QDMI calls facilitate progress monitoring, including the queue position of a job that is still waiting, and, once a terminal state is reached, retrieval of results.
    A job handle does not have to originate from a submission in the same process: given a job identifier, a session can resolve an existing circuit job into a QDMI handle and then monitor or collect it. This matters in batch settings, where the step that submits work and the step that consumes its results need not be the same process, or even the same allocation.

    The implementation maps backend-specific status values to QDMI job states. Even though a one-to-one mapping is not always possible, as the IQM Server API has more fine-grained status values than QDMI, a semantically consistent mapping is achievable, as shown in~\autoref{tab:iqm-status-map}.
    This includes normal completion, cancellation, and failure paths.

    \begin{table}[b!]
        \caption{Status and error mapping for IQM QDMI jobs.}
        \label{tab:iqm-status-map}
        \centering
        \scriptsize
        \renewcommand{\arraystretch}{1.1}
        \setlength{\tabcolsep}{4pt}
        \rowcolors{2}{lightgray!15}{white!15}
        \begin{tabular}{
            >{\raggedright\arraybackslash}p{0.45\linewidth}
            >{\raggedright\arraybackslash}p{0.45\linewidth}
        }
            \toprule
            \textbf{QDMI Status} & \textbf{IQM Server API} \\
            \midrule
            \texttt{QDMI\_JOB\_STATUS\_SUBMITTED} & \texttt{received} \\
            \texttt{QDMI\_JOB\_STATUS\_QUEUED} & \texttt{queued}, \texttt{waiting} \\
            \texttt{QDMI\_JOB\_STATUS\_RUNNING} & \texttt{validation\_started}, \texttt{validation\_ended} \\
            \texttt{QDMI\_JOB\_STATUS\_RUNNING} & \texttt{fetch\_calibration\_started}, \texttt{fetch\_calibration\_ended} \\
            \texttt{QDMI\_JOB\_STATUS\_RUNNING} & \texttt{compilation\_started}, \texttt{compilation\_ended} \\
            \texttt{QDMI\_JOB\_STATUS\_RUNNING} & \texttt{save\_sweep\_metadata\_started}, \texttt{save\_sweep\_metadata\_ended} \\
            \texttt{QDMI\_JOB\_STATUS\_RUNNING} & \texttt{pending execution}, \texttt{pending\_execution} \\
            \texttt{QDMI\_JOB\_STATUS\_RUNNING} & \texttt{execution\_started}, \texttt{execution\_ended} \\
            \texttt{QDMI\_JOB\_STATUS\_RUNNING} & \texttt{post\_processing\_pending}, \texttt{post\_processing\_started}, \texttt{post\_processing\_ended} \\
            \texttt{QDMI\_JOB\_STATUS\_RUNNING} & \texttt{running}, \texttt{processing}, \texttt{accepted}, \texttt{pending compilation}, \texttt{compiled} \\
            \texttt{QDMI\_JOB\_STATUS\_DONE} & \texttt{ready}, \texttt{completed} \\
            \texttt{QDMI\_JOB\_STATUS\_CANCELED} & \texttt{aborted}, \texttt{cancelled} \\
            \texttt{QDMI\_JOB\_STATUS\_FAILED} & \texttt{failed} \\
            \bottomrule
        \end{tabular}
    \end{table}

    Program data and execution options are passed through typed QDMI job parameters. When backend-specific options are needed, they are carried through extension fields, as demonstrated in \autoref{ex:custom-job}, rather than leaking proprietary structures into upper layers. This keeps the common path clean while still allowing advanced backend features.
    These custom parameters act as deliberate escape hatches within the QDMI specification to handle vendor-specific options (\eg, superconducting-specific heralding settings or trapped-ion shuttling configurations) without breaking the standard API footprint. As other hardware vendors adopt QDMI, they can utilize these parameters to fit their hardware blueprints. Under QDMI's governance model, a custom parameter is promoted to a standard core parameter once multiple backend implementations share the same underlying capability, ensuring consensus-driven standardization.

    Observability is handled on a separate axis from this status mapping. The mapping in \autoref{tab:iqm-status-map} is deliberately coarse: a portable scheduler cannot depend on states such as \texttt{save\_sweep\_metadata\_ended} that exist on a single backend only. Coarse status does not mean lost progress information, however, because queue visibility is carried by dedicated properties rather than by the granularity of status. As listed in \autoref{tab:parameters}, \texttt{QUEUELENGTH} reports the depth of the selected backend's queue at the device level, and \texttt{QUEUEPOSITION} reports the number of jobs ahead of a given job at the job level.

    Both properties are resolved against the IQM Server at query time rather than served from session state. Every \texttt{QUEUEPOSITION} query first refreshes the job status (cf.\ \autoref{fig:job-sequence}), so the reported position and the reported status are always mutually consistent; a stale position that no longer matches the job's state can therefore never be observed. The property is defined only while a job is queued and reports \texttt{QDMI\_ERROR\_BADSTATE} once the refreshed status has moved on, which gives a polling scheduler an unambiguous signal that waiting has ended. Where a backend supplies no trustworthy position, or does not expose queue availability at all, the corresponding query reports \texttt{QDMI\_ERROR\_NOTSUPPORTED} rather than a fabricated number.

    This placement is deliberate. Queue depth and queue position are scheduling concepts shared by virtually every queued backend, not IQM-specific ones, which is why they belong in the QDMI core rather than behind the custom-property escape hatch. The custom-parameter mechanism of \autoref{ex:custom-job} remains available for options that genuinely are vendor-specific, and the same escape hatch exists on the query side, where QDMI reserves custom device and job properties alongside the custom job parameters discussed above.

    \begin{example}\label{ex:custom-job}
    The quantum program itself, its format, and the number of shots are passed through the standardized QDMI parameters as shown in \autoref{fig:job-sequence}.
    Additionally, the QDMI device implementation supports IQM-specific execution options, which can be attached via extension fields.
    For instance, a client may request heralding mode \texttt{"zeros"} via the job parameter \texttt{CUSTOM1}, or a logical-to-physical qubit mapping such as \texttt{"q0:QB1,q1:QB2"} via \texttt{CUSTOM5}.
    During submission, these values are then translated into IQM Server API fields such as \texttt{heralding\_mode}, and \texttt{qubit\_mapping}.
    \end{example}

    Furthermore, the implementation supports both blocking and polling-style usage.
    A blocking call via \texttt{QDMI\_device\_job\_wait()} is convenient in simple scripts, where execution should pause until a job has finished.
    By contrast, polling via \texttt{QDMI\_device\_job\_check()} is more useful in orchestration settings, where many tasks may need to be tracked simultaneously.
    Internally, both operations follow the same status-check path: \texttt{wait} repeatedly invokes \texttt{check()}, which performs the backend status query and maps native IQM status values to QDMI job states.

    Notably, the job functions support both circuit execution jobs and calibration jobs (depending on whether the selected program format is \texttt{CALIBRATION} or a circuit format such as \texttt{IQMJSON}), which are mapped to distinct backend endpoints in the background.

    Result retrieval supports both aggregated measurement counts and, where available, individual shot data.
    If shot data has already been fetched for a job, histogram counts are derived locally without issuing an additional backend request.
    For calibration jobs, the first uncached result access rereads the calibration-job status, extracts the new calibration-set identifier, refreshes the session's dynamic architecture and calibration metrics to that calibration set, or retries that refresh once after 120\,s if the first attempt fails.

    \section{HPC Integration with QDMI}\label{sec:hpc-components}
    With a working device implementation in place, the question shifts to: how do existing HPC and quantum software layers connect to QDMI?
    As sketched in \autoref{fig:qdmi-hourglass}, everything above the QDMI boundary can, in principle, be shared across backends. Everything below remains vendor-specific. This section describes three areas where that reuse is most concrete: language bindings, a Qiskit-facing frontend adapter, and Slurm-based scheduler integration.

    \subsection{Language Interoperability}
    QDMI is defined as a C interface, which makes it straightforward to call from C and C++ code. For higher-level languages, however, direct C interop is cumbersome. We therefore provide idiomatic C++ and Python bindings that wrap the C entry points, manage object lifetimes, and expose QDMI sessions, queries, and jobs through native language constructs.

    The C++ layer maps QDMI handles to RAII-managed objects and translates error codes into exceptions where appropriate. The Python layer builds on the C++ bindings via nanobind.
    We note that while a low-level standard like QDMI does not strictly require high-level bindings (since Python can natively interface with C APIs via standard foreign function interfaces like \texttt{ctypes}), providing official pre-packaged Python wrappers simplifies integration. Early ecosystem feedback indicated that maintaining native-feeling bindings would significantly lower the barrier to entry and prevent redundant, manual binding implementations across different frontend adapters (such as Qiskit or PennyLane).

    These bindings are not IQM-specific. Any QDMI device plugin, regardless of vendor, can be loaded and used through the same C++ or Python API. In practice, this means that workflow code written using the Python bindings works even when the underlying device changes from IQM to another backend.

    \subsection{Frontend Integration: Qiskit over QDMI}
\begin{figure}[t]
        \centering
        \includegraphics[width=\columnwidth]{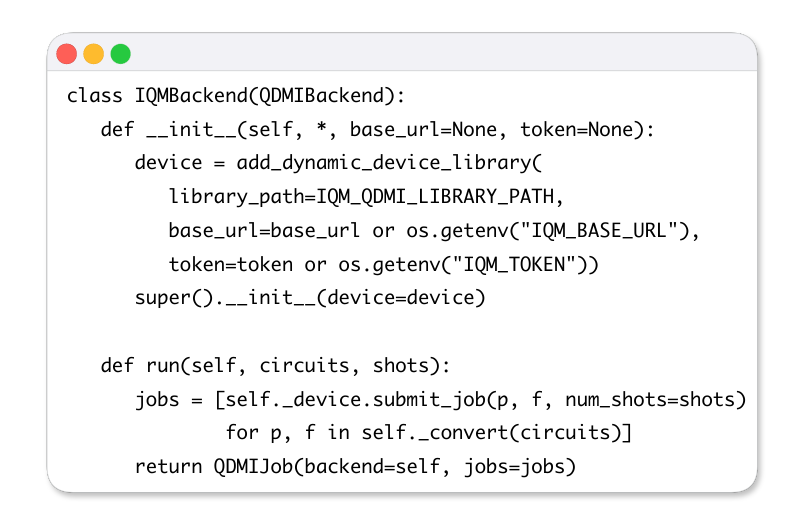}\vspace*{-5mm}
        \caption{Qiskit-facing adapter skeleton over QDMI.}
        \label{fig:qiskit-adapter}
    \end{figure}

    The Qiskit adapter builds on the Python bindings and implements Qiskit's \texttt{BackendV2} interface on top of QDMI (cf. \autoref{fig:qiskit-adapter}). Concretely, property queries during transpilation target construction, circuit submission, and result retrieval all map to QDMI session and job calls rather than to any vendor-specific client library.

    Beyond the backend interface, the adapter also provides \texttt{SamplerV2} and \texttt{EstimatorV2} implementations of Qiskit's primitive abstractions. Because both primitives delegate to QDMI internally, they work with any QDMI device, not only with IQM hardware. From a user's perspective, Qiskit code requires no modification: circuits are transpiled, executed, and measured through the usual Qiskit workflow. The difference is that the backend can now also be an HPC-accessible quantum computer interfaced via QDMI, rather than only a cloud endpoint managed by a vendor-specific provider package.

    This design consolidates adapter maintenance at the QDMI boundary. Without it, every hardware vendor would need to ship and maintain its own Qiskit provider, and every Qiskit API change would need to be propagated independently across those providers. With a single QDMI-targeting adapter, frontend evolution and backend evolution are decoupled.

    While our case study focuses on a Qiskit workflow, the same pattern applies directly to other popular frontend frameworks, such as PennyLane~\cite{bergholm2018pennylane} and CUDA-Q~\cite{nvqlink2025}. For instance, a dedicated PennyLane-QDMI device plugin can map PennyLane's execution tapes to QDMI sessions and jobs using the Python bindings. Similarly, since QDMI is a native C/C++ library, CUDA-Q's compiler and runtime stack can link directly against it, executing circuits and querying calibration properties without any Python runtime overhead. The primary barrier to adoption for these frameworks is vendor-side support: frameworks can only target QDMI if backend vendors implement the corresponding device-level plugin (such as our implementation for IQM). Additionally, mapping framework-specific compiler hints and error mitigation settings through QDMI's custom parameters requires standardization across different frontend toolchains.

    \subsection{Scheduler Integration in Practice}\label{sec:hpc-components:scheduler}
    From a user's perspective, moving from interactive QDMI usage to Slurm-dispatched execution requires few changes. The application code is the same; only the launch method differs. Users write a standard \texttt{sbatch} script, and quantum calls within the application use the same QDMI session and job path as in an interactive shell.

    The more interesting design question is on the operator side: how and where are endpoint URLs and credentials injected? We considered two approaches. The first is script-level configuration, in which each job script explicitly sets the required environment variables. This works well for exploratory testing and early onboarding. The second is centralized injection through a SPANK plugin that runs inside the Slurm prolog. In this mode, the plugin reads a center-level configuration file, resolves the correct IQM endpoint and token path for the allocated partition, and exports the corresponding environment variables before the user's application starts. Users never directly touch endpoint URLs or credential paths.

    Within the SPANK-based approach, the plugin handles credentials in one of two modes, depending on the center's security requirements. In \emph{User Mode}, the plugin securely reads a user's personal configuration or API token from their home directory (\eg, \texttt{\textasciitilde/.iqm/tokens.json}) and exports the path to the user's environment. In \emph{Operator Mode}, designed for shared on-premise partitions, the plugin injects a partition-wide, operator-managed backend credential directly before launch. This ensures that users can execute scheduled jobs without needing direct access to backend administrative credentials or manual key generation.

    The SPANK-based approach has two practical advantages. First, credential rotation and endpoint changes become operator-level configuration updates rather than user-facing changes. Second, the plugin can enforce center policies, \eg, restricting which partitions may access which backends, or verifying that the token file is valid before the job starts. Both reduce the support burden that comes with manual environment setup across many users.
    At the same time, it has the downside of relying on SPANK as a plugin system, which might be prohibitive for some HPC centers.

\begin{figure}[t]
        \centering
        \includegraphics[width=\columnwidth]{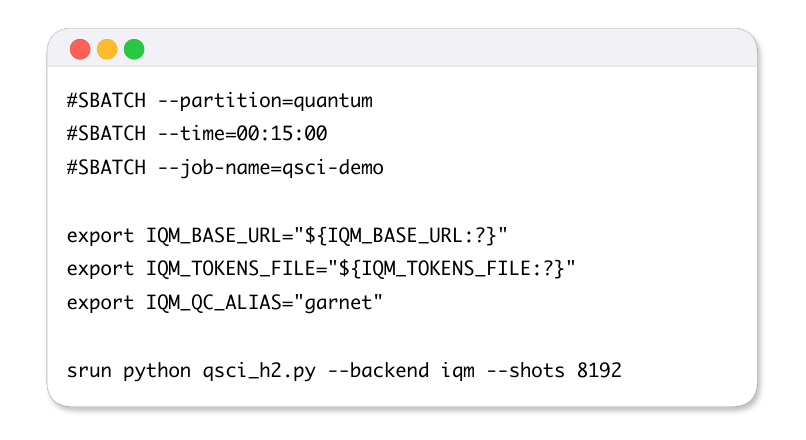}\vspace*{-4mm}
        \caption{Illustrative Slurm submission pattern for QDMI-enabled execution.}
        \label{fig:slurm-qdmi-submit}\vspace*{-2mm}
    \end{figure}

    \autoref{fig:slurm-qdmi-submit} shows a minimal submission script. The user specifies only Slurm resource parameters and the application command; all QDMI-related configuration is injected automatically.

    Beyond raw \texttt{sbatch}/\texttt{srun} scripting, we provide a thin Python convenience layer, the \emph{offloader}, that packages this dispatch pattern into two calls, \texttt{sample()} and \texttt{estimate()}, matching the sampling and estimation primitives in \autoref{sec:hpc-components}. Each call accepts a circuit (and, for \texttt{estimate()}, an operator and an ansatz), serializes it to shared storage, and invokes \texttt{srun} on the quantum partition, where a small companion executable deserializes the payload, runs it through the Qiskit-facing adapter, and returns the result. A \texttt{local} flag switches between this Slurm dispatch and in-process execution against the same QDMI-backed primitives, and a \texttt{simulator} flag independently switches the target between the QPU and a classical simulator backend; an optional quantum-computer selector is passed through as an \texttt{srun} option that the SPANK plugin resolves into the job's environment, so a single call site can target a specific device without hardcoding endpoint details. Because both flags are simple booleans (or a device selector) rather than separate code paths, the same call expresses all four combinations of local/Slurm-dispatched and simulator/hardware execution, which is exactly the toggle pattern used by the end-to-end demonstration in \autoref{sec:demo}. A \texttt{licenses} argument is forwarded to \texttt{srun} in the same manner. Because the device queues concurrent submissions itself (cf.\ \autoref{sec:prelim}), nothing about correctness depends on this argument, and omitting it is a no-op; its purpose is to let a center tune the number of concurrent calls against the shared quantum backend, that is, to regulate the outer of the two queues, trading queue wait time against classical- and quantum-resource idle time according to local policy. This matters mainly on shared multi-tenant clusters that front single-tenant on-premise hardware.

    \section{Deployment Scenarios and Tradeoffs}\label{sec:deployment}
    Across the components described in the previous section, user code and adapter logic remain the same regardless of how a center deploys its quantum resources; only endpoint and credential configuration change. This consistency is the direct payoff of standardizing at the QDMI boundary and motivates the deployment styles we evaluate here.

    \subsection{Cloud QPU}
    The first style is loose coupling: classical computation runs in a standard environment, while quantum execution is routed to a remote (cloud) backend via QDMI. This is often the fastest way to get started and is useful for early user onboarding. This approach remains valuable for interactive development, allowing end users to craft hardware-specific algorithms, improve the compilation and optimization performance of those algorithms, and conduct full-pipeline testing prior to production.

    Much of the early work followed this pattern without unified resource orchestration. Quantum computers were accessed via cloud, and the resulting data was uploaded to a separate, disconnected HPC center~\cite{sriluckshmy2025ggut}. Running iterative loops with this disconnection is difficult, and workflows were typically executed sequentially without feedback from earlier steps. In practice, this means either blocking classical and quantum resources for the entire workflow duration or releasing them for indefinite periods. For a quantum computer, releasing the resource returns the job to the general-purpose queue inside the QPU, increasing the wall-clock time.

\begin{figure}[t]
        \centering
        \includegraphics[width=\columnwidth]{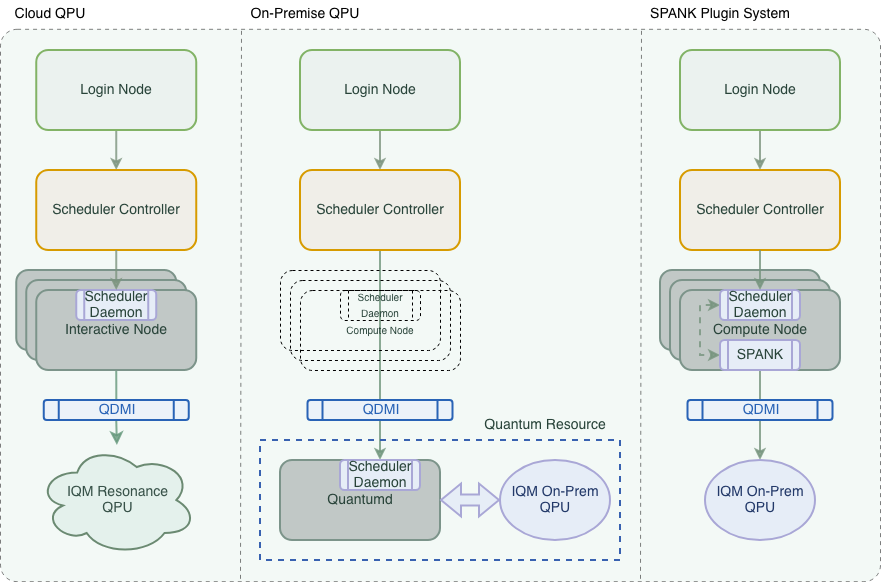}
        \caption{Deployment scenarios enabled by the same QDMI-based integration path.}\vspace*{-2mm}
        \label{fig:deployment-modes}
        \vspace{-1\baselineskip}
    \end{figure}

    Using this deployment scenario, illustrated in \autoref{fig:deployment-modes}, allows users to retain full internal accountability and traceability over system resources. Furthermore, leveraging existing local infrastructure to interface with on-premises or cloud-based quantum resources is cost-effective because it avoids additional authorization software layers. Operationally, this strategy ensures that standard cluster tools remain available for troubleshooting, significantly streamlining user support. We note one deployment constraint that the cloud style inherits from center security policy: compute nodes are frequently barred from outbound internet access, so a remote backend may be reachable only from login or gateway nodes or through a center-approved proxy. 
    Establishing a clean QDMI boundary can facilitate the refinement of interconnection interfaces.

    \subsection{On-Premise QPU}
    The second style is tight coupling with scheduler-managed quantum access in a working-node mode. Users stay within the center workflow, submit quantum jobs through the same scheduling and accounting paths as other workloads, and provide the required backend configuration from their usual HPC job environment. This lets the HPC center co-schedule classical and quantum resources, reducing idle time on scarce QPU hardware. Shirakawa~\emph{et al.}~\cite{shirakawa2025closedloop} demonstrated this style in practice: the QPU samples distributions representing electron orbitals in parallel with classical nodes computing new variational parameters, closing the optimization loop without leaving the scheduler.

    This style is operationally heavier, requiring partition configuration, accounting integration, and credential management. However, it aligns best with production policies and enables iterative hybrid workflows that loose coupling cannot support efficiently. At the same time, because the majority of the components required for the system's proper operation are already part of the HPC center, this results in a robust execution platform in which a QPU can be connected with minimal interference.

    This model fits batch-oriented systems and supports iterative hybrid jobs, but more complex workflows that coordinate multiple HPC and quantum resources will likely require richer orchestration mechanisms.

    The same scheduler-managed model is also a natural starting point for future workflows that coordinate multiple QPUs or combine classical and quantum resources under a single orchestration layer, although such scenarios are beyond the scope of this paper.

    \subsection{SPANK Plugin Integration}
    The third style extends tight coupling with scheduler-side automation. Here, a SPANK plugin injects configuration and enforces guardrails at job launch time, as described in \autoref{sec:hpc-components:scheduler}. By relying on automation mechanisms already standard in HPC centers, this approach extends established operational practices to quantum workloads rather than introducing a separate control path.
    This eliminates repetitive user-side setup, gives operators direct control over endpoint selection, credential management, and access policies (supporting both User Mode and Operator Mode credential injection), and makes it easier to incorporate quantum jobs into existing workflow managers and center orchestration tools.

    This scheduler-level automation enables user workflows to scale while maintaining operator control over guardrails, ensuring proper use of this shared system. Consequently, this capability enables seamless scaling to a larger number of on-premises quantum computers and lets classical and quantum work overlap: the device's own FIFO queue absorbs concurrent submissions, while a Slurm license lets a center bound how many jobs queue up at once. Quantifying the utilization gain that follows from this overlap is beyond the scope of this integration study and is left to future work, together with the scheduling policies that would build on it.

    \section{Demonstration: End-to-End QSCI Workflow}\label{sec:demo}

    One of the most promising near-term workflows that combines quantum and HPC resources is \emph{Quantum Selected Configuration Interaction}~(QSCI)~\cite{Kanno2023,Moreno2025}. In the quantum chemistry setting, QSCI approximates the low-lying spectrum of a many-electron Hamiltonian by using quantum samples to identify a reduced configuration subspace and then diagonalizing the corresponding truncated Hamiltonian classically. The workflow starts with a molecular problem definition, maps the second-quantized Hamiltonian to qubits, prepares a parameterized trial state, samples it on a simulator or QPU, selects the dominant bitstring configurations consistent with the target particle-number sector, and finally solves the reduced eigenproblem on classical resources. The key systems point is that this sequence naturally splits across heterogeneous resources: chemistry preprocessing and orchestration on the login node, quantum execution on the QPU-facing path, and reduced-subspace construction and diagonalization on classical compute nodes.

    QSCI is therefore a natural use case to showcase the workflow flexibility enabled by QDMI for two reasons. First, it combines tightly coupled classical and quantum stages rather than submitting a single isolated circuit. In our implementation, the initial state is prepared using a Hartree--Fock reference and a UCCSD ansatz; the variational parameters are optimized with VQE; the optimized state is sampled; and the resulting counts drive a reduced-space diagonalization. This is representative of the orchestration pattern relevant to HPC centers: the quantum step is only one component of a larger workflow whose control logic, data movement, and post-processing remain classical.

    \begin{figure}
        \centering
        \includegraphics[width=0.99\columnwidth]{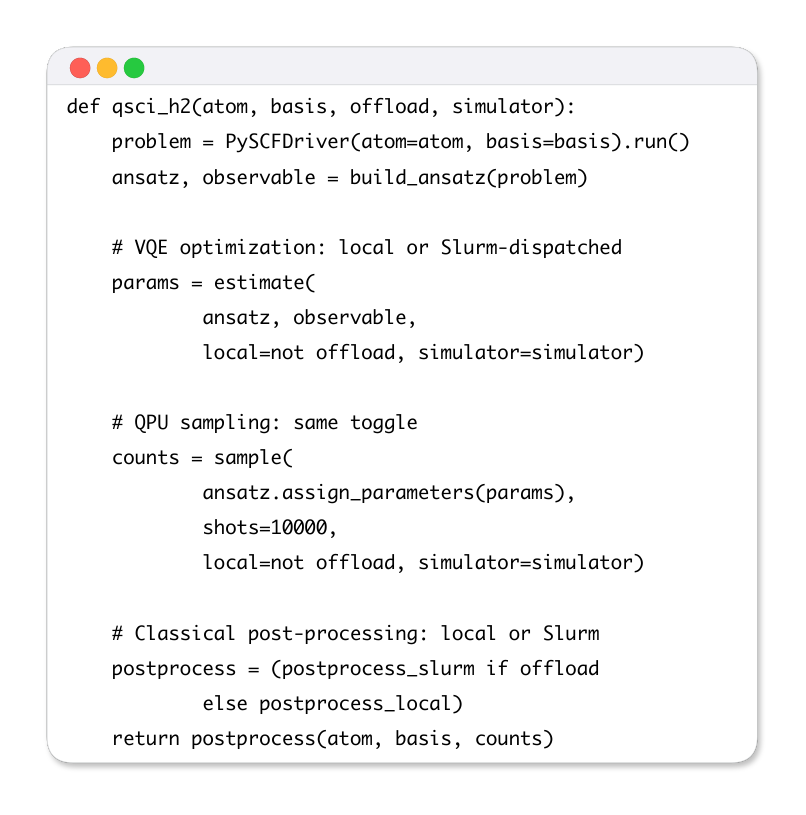}
        \vspace{-0\baselineskip}
        \caption{Simplified QSCI control flow showing execution-mode toggling. The \texttt{offload} and \texttt{simulator} flags are the only parameters that differ between local simulation, Slurm-dispatched simulation, local hardware, and Slurm-dispatched hardware execution.}
        \label{fig:qsci-toggle}
    \end{figure}

    Second, the same workflow must be able to target multiple execution environments. During development, users want to validate the quantum path locally against a simulator; during integration, they may offload the same computation to a cloud QPU; and in production, they may direct it to an on-premise QPU endpoint. Independently, the classical post-processing may remain local for small molecules or be dispatched to cluster resources for larger workloads. QDMI makes the application-side code largely independent of this deployment mode. The algorithm code and adapter wiring can, therefore, be reused across local development, scheduled execution, and hardware-backed runs.

    This makes the benefit of our demonstration apparent. We are not looking to demonstrate a benchmark number; rather, we demonstrate a single integration path that allows seamless integration of local execution, Slurm-dispatched execution, and real hardware access. The execution mode is toggled in the application code, while the endpoint and credential contexts are injected at runtime via the backend configuration and authentication environment. In our implementation, this context includes the IQM endpoint, the authentication token or token file, and an optional quantum-computer selector, but the QSCI algorithm itself remains unchanged.

    \autoref{fig:qsci-toggle} details a corresponding QSCI usage example. The \texttt{offload} flag selects between local execution and Slurm-dispatched execution; the \texttt{simulator} flag selects between a classical simulator and the real QPU. Both the VQE parameter-estimation step and the sampling step use the same pair of flags, so switching from a local simulation run to a Slurm-dispatched hardware run requires changing two Boolean arguments and nothing else. The algorithm code, the QDMI session path, and the adapter integration path remain identical across all four configurations.

\begin{figure}
        \centering
        \includegraphics[width=0.8\columnwidth]{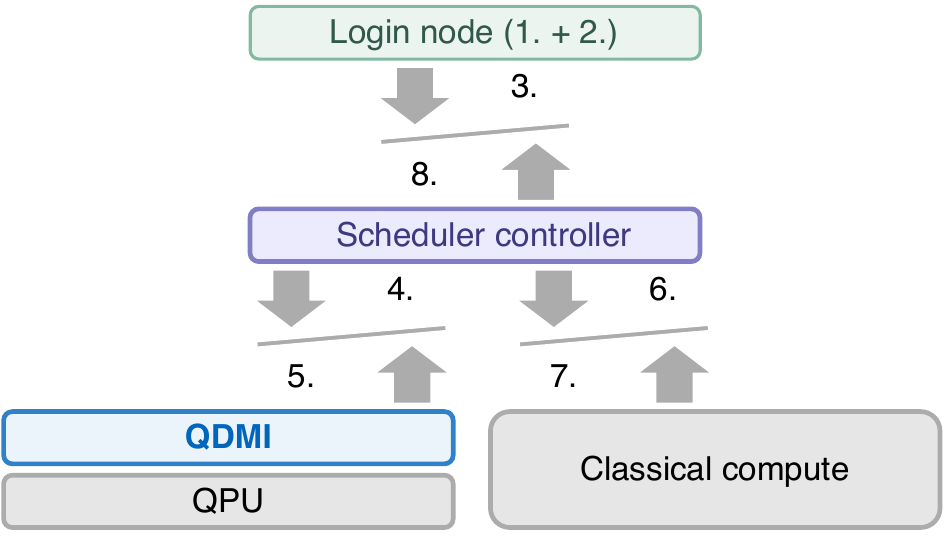}
        \caption{Schematic end-to-end workflow for QSCI execution. Numbers represent the workflow's sequence of operations and are explained in the main text. QDMI facilitates the communication in steps 4 and 5.}
        \label{fig:qsci-sequence}
        \vspace{-1.5\baselineskip}
    \end{figure}

    Our QSCI workflow is shown in Fig.~\ref{fig:qsci-sequence}. The numbers in the figure represent steps of the algorithm as follows:
    \begin{enumerate}
        \item The molecular problem is formulated, typically on the login node. For the QSCI workflow, this means building the problem and deriving the corresponding qubit Hamiltonian.
        \item A parameterized input state is constructed. In our implementation, this uses a Hartree--Fock initial state and a UCCSD ansatz, followed by a VQE optimization loop to obtain circuit parameters.
        \item If offloading is enabled, the ansatz and operator are serialized on shared storage and submitted through Slurm to the quantum partition; otherwise, the same logic executes locally in process.
        \item On the quantum path, the Qiskit-facing adapter invokes the QDMI-backed IQM device, which submits the circuit to either a simulator or the target QPU depending on the runtime context.
        \item The optimized circuit is sampled, and the resulting bitstring counts are returned through the same adapter and scheduler path.
        \item The sampled configurations define the reduced QSCI basis. Classical post-processing constructs the truncated Hamiltonian in that basis and either diagonalizes it locally or performs Slurm-dispatched classical execution.
        \item The approximate eigenvalue(s) and eigenvector(s) are returned to the workflow controller.
        \item The application returns the QSCI result to the user while preserving the same execution contract across all deployment modes.
    \end{enumerate}

    In our setup, classical preparation runs in Python on the login node. Quantum execution requests go through the Qiskit-facing adapter, into QDMI bindings, through the IQM QDMI device, and finally to the IQM backend. The returned results follow the same path back to the application for optimization and post-processing. The scientific claim of this section is therefore modest but important for HPC practice: not that this example sets a chemistry benchmark, but that one execution contract spans local development, scheduler-mediated execution, and real hardware access without rewriting the application-level workflow. The complete implementation of the QSCI proof-of-concept is available in the \texttt{examples/} directory of our public repository, and detailed execution instructions for various test cases can be found in the documentation at \url{https://iqm-finland.github.io/QDMI-on-IQM/examples.html}.

    \section{Conclusion}\label{sec:conclusion}
    This paper examined a practical HPCQC integration path built around a standardized device-management boundary. Using IQM systems as a real-hardware case study, we implemented an IQM-backed QDMI device layer and connected it to Slurm execution and Qiskit-facing frontend workflows. The implementation is publicly available at \href{https://github.com/iqm-finland/QDMI-on-IQM}{github.com/iqm-finland/QDMI-on-IQM}.

    The central result is straightforward. When the device-management boundary is standardized, software above that boundary can be reused across deployment settings. In our implementation, this includes language bindings, the Qiskit-facing adapter, the Slurm integration pattern, and the associated deployment paths.

    The end-to-end QSCI study provides operational evidence for this point. The same application-side path was used for local simulation, Slurm-dispatched simulation, local hardware access, and Slurm-dispatched hardware access; only execution mode and backend configuration changed. The contribution is therefore not a chemistry performance result but an integration result: a single interface boundary can support multiple execution contexts without rewriting the workflow.

    This does not remove all HPCQC integration work. Backend-specific payloads, calibration behavior, and operational constraints remain. Our result is narrower and more practical: these differences can be contained within the device-management boundary rather than propagating through user code, frontend tooling, and scheduler integration.

    For HPC centers, this yields deployment flexibility rather than a single prescribed model: cloud access for early exploration, scheduler-managed on-premise access in working-node mode, and SPANK-automated access when tighter operator control is required. The application-side software path stays the same across these options.

    Overall, the case study indicates that device-management standardization is already feasible on production hardware and useful in day-to-day integration work. It offers centers, vendors, and software maintainers a common integration point without forcing one deployment style. Future work can build on this boundary to study broader scheduling policies and multi-resource orchestration.

    \section*{Acknowledgments}
    While preparing the manuscript, Claude Sonnet 4.6, Claude Opus 4.6, Gemini 3 Pro, and GPT 5.3 Codex were used via GitHub Copilot to improve readability, spelling, grammar, and clarity throughout.
    Each LLM output was reviewed by the authors and edited manually as needed.
    The authors take full responsibility for the final content.

    L.B., M.W., P.H., and R.W. acknowledge funding from the ERC under the EU's Horizon 2020 research and innovation program (No. 101001318 and No. 101114305), the Munich Quantum Valley, supported by the Bavarian state government with funds from the Hightech Agenda Bayern Plus, and the BMFTR under 13N17298 (SYNQ) and 01MQ25001I (FullStaQD), and the DFG under 563402549 and 563436708.

    \balance
    \printbibliography

\end{document}